\documentclass[aps,pra,10pt,amsmath,notitlepage,nofootinbib,twocolumn,superscriptaddress]{revtex4-2}
\pdfoutput=1
\usepackage[utf8]{inputenc}
\usepackage[english]{babel}
\usepackage[T1]{fontenc}
\usepackage{microtype}

\usepackage{amsfonts}
\usepackage{amsmath}
\usepackage{amssymb}
\usepackage{graphicx}
\usepackage[dvipsnames]{xcolor}

\usepackage[breaklinks=true,colorlinks=true,linkcolor=teal,urlcolor=teal,citecolor=teal]{hyperref}
\usepackage{orcidlink}

\let\originalleft\left
\let\originalright\right
\renewcommand{\left}{\mathopen{}\mathclose\bgroup\originalleft}
\renewcommand{\right}{\aftergroup\egroup\originalright}

\DeclareMathOperator{\Tr}{Tr}

\let\Re\relax

\DeclareMathOperator{\Re}{Re}

\newcommand{\nth}{\bar{n}_{th}}

\begin{document}

\title{Homodyne Unravelling and Parameter Estimation in Squeezed Thermal Baths}

\author{Eoin O'Connor\,\orcidlink{0000-0002-8900-6649}}
\email{eoin.oconnor@unimi.it}
\affiliation{Dipartimento di Fisica {\it "Aldo Pontremoli"}, Universit\`a degli Studi di Milano, I-20133 Milano, Italia}

\author{Francesco Albarelli\,\orcidlink{0000-0001-5775-168X}}
\email{francesco.albarelli@gmail.com}
\affiliation{Università di Parma, Dipartimento di Scienze Matematiche, Fisiche e Informatiche, I-43124 Parma, Italy}
\affiliation{INFN—Sezione di Milano-Bicocca, Gruppo Collegato di Parma, I-43124 Parma, Italy}

\author{Marco G. Genoni\,\orcidlink{0000-0001-7270-4742}}
\email{marco.genoni@unimi.it}
\affiliation{Dipartimento di Fisica {\it "Aldo Pontremoli"}, Universit\`a degli Studi di Milano, I-20133 Milano, Italia}

\begin{abstract}
We derive an alternative formulation of the stochastic master equation describing the evolution of quantum systems continuously measured via homodyne detection in a squeezed-thermal environment. This formulation allows us to identify two distinct collapse operators, corresponding respectively to the monitored and unmonitored channels, and to express the quantum state evolution directly in terms of Kraus-like operators. As a consequence, it facilitates the application of advanced numerical methods that account for higher-order corrections and discrete-time monitoring, and it also enables the efficient numerical evaluation of Cramér-Rao bounds in parameter estimation problems.
To demonstrate the latter, we evaluate the Fisher information of the continuous homodyne monitoring for qubit frequency estimation in the presence of a squeezed-thermal bath. Our results show that exponential scaling of the steady-state Fisher information rate with squeezing is retained even at non-zero temperature.
\end{abstract}

\maketitle

\section{Introduction}
Continuously monitored quantum systems are of fundamental importance to the conceptual and theoretical frameworks of quantum state engineering protocols and quantum estimation~\cite{WisemanMilburn,jacobs2014quantum,JordanSiddiqi,AlbarelliPLA2024}, and have recently been demonstrated experimentally in several platforms including superconducting circuits~\cite{Murch2013Observing,CampagneIbarcq2016Observing,ficheux2018dynamics,Minev2019Catch}, optomechanical systems~\cite{wieczorekOptimalStateEstimation2015,rossiObservingVerifyingQuantum2019}, and atomic systems~\cite{duan_concurrent_2025}. 

The evolution of the density operator for a continuously monitored quantum system is governed by stochastic master equations (SMEs), which are typically referred to as unravellings of the corresponding unconditional Markovian master equation (ME)~\cite{AlbarelliPLA2024}. There are two main types of unravellings: jump-like and diffusive, physically corresponding to continuous photodetection and continuous homodyne (or general-dyne) detection, respectively.

In this manuscript we focus on the dynamics of quantum systems continuously monitored via homodyne detection in the presence of a squeezed-thermal bath. This scenario is particularly relevant in superconducting circuit-QED experiments: residual thermal microwave photons in the readout resonator can remain a significant source of noise even at dilution-refrigerator temperatures~\cite{Yan2018,Wang2019}. Therefore, finite-temperature effects in continuous dispersive readout have to be incorporated into the stochastic description of circuit-QED measurements~\cite{Cernotik2015}.
The same reasoning applies to qubit readouts via a microwave resonator prepared in a squeezed (thermal) state~\cite{Jakob2001,Mundarain2006,Didier2015,Eddins2018,Dassonneville2021}.
In this regard, phase-dependent inhibition of the decay of atomic coherence in a squeezed environment was predicted by Gardiner~\cite{Gardiner1986} and demonstrated experimentally with a superconducting artificial atom~\cite{Murch2013a}.
To our knowledge, conditional quantum trajectories of an atom coupled to a squeezed environment have not yet been reconstructed experimentally from continuous homodyne records, although measurements of the fluorescence spectrum have been performed in this setup~\cite{Toyli2016}.

Our main result is the derivation of an alternative form of the SME describing the homodyne unravelling in a squeezed-thermal environment. 
The benefit of our derived equation is that it immediately identifies the evolution in terms of two Lindblad jump operators, corresponding to the monitored and unmonitored channels, respectively. 
Related results have appeared in a more formal treatment via quantum stochastic calculus~\cite{Nurdin2014,Gough2026a}, without presenting the SME with separate monitored and unmonitored jump operators, and without a detailed discussion of its practical applications.
In fact, as a byproduct, this new formulation enables us to express the evolution using Kraus-like operators, following the recipe described in~\cite{ROUCHON20222,Amini2011,Jordan2016}. Consequently, one can straightforwardly include higher-order corrections~\cite{Rouchon2015,Guevara2020,Wonglakhon2024} and the experimentally relevant scenario of time-averaged continuous measurements~\cite{robinet,wonglakhon2026,hutin_robinet_2026}. Furthermore, this formulation makes it possible to leverage existing numerical routines, such as those in the QuTiP package~\cite{JOHANSSON20121760,JOHANSSON20131234}, to simulate diffusive SMEs.

Such a Kraus-like representation is crucial in numerically evaluating the Cram\'er-Rao bound corresponding to continuous homodyne detection~\cite{GammelmarkMolmer,AlbarelliQuantum}.
We thus provide an example in this direction by evaluating the corresponding Fisher information for frequency estimation in a qubit via homodyne detection in a squeezed-thermal environment.
More generally, we also present a more efficient algorithm for the Fisher information of continuous homodyne measurements, which removes a zero-mean stochastic contribution present in previous approaches (see Appendix~\ref{a:FI_calc}).

The manuscript is organized as follows:  In Sec.~\ref{s:squeezed_thermal_ME_SME} we give a brief introduction to the Markovian master equation in the presence of a thermal squeezed bath and its standard unravelling for continuous homodyne detection. Section~\ref{s:equivalent_SME} derives an equivalent homodyne SME in the presence of a squeezed thermal environment. In Sec.~\ref{s:frequency_estimation} we apply our result to evaluate the Fisher information for frequency estimation via continuous homodyne detection in a squeezed-thermal bath. Sec.~\ref{s:conclusions} discusses implications and future directions.
%
\section{Markovian ME and Homodyne SME for a squeezed-thermal environment}
\label{s:squeezed_thermal_ME_SME}
\subsection{Markovian master equation for a squeezed-thermal environment}
Let us consider a quantum system interacting with a Markovian squeezed-thermal bath. Its non-unitary evolution is described by a Markovian master equation
\begin{align}
\frac{d\rho}{dt} &= \mathcal L\rho \,\\
&=
\kappa(N+1)\mathcal D[\hat{c}]\rho
+
\kappa N \mathcal D[\hat{c}^\dagger]\rho \nonumber \\
&\,\,\, \,\,\,\, +
\frac{\kappa M}{2}[\hat{c}^\dagger,[\hat{c}^\dagger,\rho]]
+
\frac{\kappa M^*}{2}[\hat{c},[\hat{c},\rho]] \,,
\label{eq:me_v1}
\end{align}
where we have introduced the Lindblad superoperator $\mathcal{D}[\hat{O}]\rho = \hat{O} \rho \hat{O}^\dag - (\hat{O}^\dag \hat{O} \rho + \rho \hat{O}^\dag\hat{O})/2$.
The parameters $N$ and $M$ are linked to the thermal and squeezing properties of the environment, via the equations:
\begin{align}
    N &= \nth \cosh^2 r + (\nth+1) \sinh^2 r \,,\\
    M &= e^{i\theta} \sinh r \cosh r (2 \nth +1) \,,
\end{align}
where $\bar{n}_{th}$ corresponds to the number of thermal excitations of the bath, while $\xi = r e^{i\theta}$ represents its squeezing parameter. By means of these parameters, one can then write the master equation~\eqref{eq:me_v1} in the Lindblad form as
\begin{align}
  \frac{d\rho}{dt} = \kappa(\nth+1)\mathcal D[\hat{b}]\rho
+
\kappa \nth \mathcal D[\hat{b}^\dagger]\rho  \,,
\label{eq:me_v2}
\end{align}
where we have defined the {\it squeezed annihilation operator} $\hat{b} = \cosh r \hat{c} - e^{i\theta}\sinh r \hat{c}^\dag$.
It is then straightforward to write the two relevant limiting cases: i) the thermal master equation, with no squeezing ($r=0$) and thus $N=\nth$, yielding
\begin{align}
    \frac{d\rho}{dt} = \kappa(\nth+1)\mathcal D[\hat c]\rho
    +
    \kappa \nth \mathcal D[\hat c^\dagger]\rho  \,,
\end{align}
and ii) the pure-squeezed bath master equation, where $\nth=0$ leading to
\begin{align}
  \frac{d\rho}{dt} = \kappa \mathcal D[\cosh r \hat{c} - e^{i\theta}\sinh r \hat{c}^\dag ]\rho \,.
\end{align} %
\subsection{Homodyne SME for a squeezed-thermal bath}
Let us now assume that the environment is continuously monitored via homodyne detection. For the sake of simplicity, we will fix the homodyne phase $\phi=0$, but the result can easily be generalised by making the transformation
\begin{equation}
    \hat c \to e^{-i \phi} \hat c, \quad M \to M e^{-2 i \phi}. 
\end{equation}
Before writing the SME describing the conditional state evolution, let us first define the {\it monitored} operator 
\begin{align}
    \hat{A}_m  = \sqrt{\frac{\kappa}{\Lambda}}
    \left[(N+M^*+1)\hat{c}-(N+M)\hat{c}^\dagger\right]\,,
    \label{eq:Am}
\end{align}
where $\Lambda := 1 + 2N + M + M^*$. The homodyne SME then reads~\cite{WisemanPhDThesis,GeneralDino,AlbarelliPLA2024}:
\begin{align}
    d\rho_c = \mathcal{L}\rho_c \,dt + \mathcal{H}[\hat{A}_m] \rho_c \,dw_t \,,
    \label{eq:SME_v1}
\end{align}
where we have introduced the measurement back-action superoperator $\mathcal{H}[O]\rho =\hat{O} \rho + \rho \hat{O}^\dag - \langle \hat{O} + \hat{O}^\dag \rangle \rho$, with $\langle \hat{O} \rangle = \Tr[\rho \hat{O}]$. The Wiener increment $dw_t$ is related to the measured photocurrent via the equation
\begin{align}
    dy_t = I(t) dt = \langle \hat{A}_m + \hat{A}_m^\dagger \rangle \,dt + dw_t \,.
    \label{eq:photocurrent}
\end{align}
It is easy to check that the monitoring opereator $\hat{A}_m$ entering in the backaction superoperator $\mathcal{H}[\hat{A}_m]$ does not correspond to either of the two collapse operators $\hat{d}_1 = \sqrt{\kappa (\nth +1)}\hat{b}$ and $\hat{d}_2 =\sqrt{\kappa \nth} \hat{b}^\dagger$ appearing in the unconditional Lindlbad master equation~\eqref{eq:me_v2} (except for the pure-squeezed bath $\nth =0$ that we will better discuss later in the manuscript). We remind that in fact, one could write a valid diffusive SME of the form
\begin{align}
    d\rho_c &= \mathcal{D}[\hat{d}_1]\rho_c \,dt + \mathcal{D}[\hat{d}_2]\rho_c \,dt \nonumber \\
    &\,\,\,\,\,\, \mathcal{H}[\hat{d}_1]\rho_c \,dw_t^{(1)} +
    \mathcal{H}[\hat{d}_2]\rho_c \,dw_t^{(2)}
\end{align}
corresponding to two photocurrents $dy_t^{(j)} = \langle \hat{d}_j + \hat{d}_j^\dagger \rangle \,dt + dw_t^{(j)}$ with $j\in \{1,2\}$. However, one can show how this unravelling corresponds to a very idealized (and impractical) scenario where the output fields corresponding to the two-mode purification of the squeezed-thermal bath are jointly monitored via homodyne detection~\cite{Ravotto}.

\section{An equivalent homodyne SME for a squeezed-thermal environment}
\label{s:equivalent_SME}
Our goal in this section is to derive a homodyne SME equivalent to Eq.~\eqref{eq:SME_v1}, with a clear distinction between monitored and unmonitored dynamics. Specifically, we aim to derive the superoperator $\mathcal{L}_u$, such that we can write
\begin{align}
    d\rho_c &= \mathcal{L}_u \rho_c \,dt + \mathcal{D}[\hat{A}_m]\rho_c \,dt \nonumber \\
    &\,\,\,\,\,\, + \mathcal{H}[\hat{A}_m]\rho_c \,dw_t \,.
    \label{eq:sme_v2}
\end{align}
By comparing Eq.~\eqref{eq:SME_v1} and Eq.~\eqref{eq:sme_v2}, we are thus defining the residual {\it unmonitored} generator
\begin{align}
    \mathcal{L}_u = \mathcal{L} - \mathcal{D}[\hat{A}_m] \,.
    \label{eq:Lu}
\end{align}
In order to evaluate it, let us use the identity
\begin{align}
    \mathcal D[\alpha \hat{c} + \beta \hat{c}^\dagger]\rho
    &=
    |\alpha|^2 \mathcal D[\hat{c}]\rho
    +
    |\beta|^2 \mathcal D[\hat{c}^\dagger]\rho \nonumber \\
    &\,\,\,\,\,
    -
    \frac{\alpha\beta^*}{2}[\hat{c},[\hat{c},\rho]]
    -
    \frac{\alpha^*\beta}{2}[\hat{c}^\dagger,[\hat{c}^\dagger,\rho]] \,.
\label{eq:identity}
\end{align}
By defining 
\begin{align}
    \alpha_{\rm m} &=
\sqrt{\frac{\kappa}{\Lambda}}(N+M^*+1), \\
\beta_{\rm m}
&=
-\sqrt{\frac{\kappa}{\Lambda}}(N+M),
\end{align}
we can exploit the identity above to write
\begin{align}
    \mathcal D[\hat{A}_m]\rho
    &=
    |\alpha_{\rm m}|^2 \mathcal D[\hat{c}]\rho
    +
    |\beta_{\rm m}|^2 \mathcal D[\hat{c}^\dagger]\rho \nonumber \\
    &\,\,\,\,\,
    -
    \frac{\alpha_{\rm m}\beta_{\rm m}^*}{2}[\hat{c},[\hat{c},\rho]]
    -
    \frac{\alpha_{\rm m}^*\beta_{\rm m}}{2}[\hat{c}^\dagger,[\hat{c}^\dagger,\rho]] \,.
\end{align}
Subtracting the coefficients $\mathcal{D}[\hat{c}]\rho$ gives
\begin{align}
    \kappa(N+1)-|\alpha_{\rm m}|^2
    =
    \frac{\kappa\big(N(N+1)-|M|^2\big)}{\Lambda} \,.
\end{align}
Similarly, by subtracting the terms multiplying $\mathcal{D}[\hat{c}^\dagger]$ we obtain
\begin{align}
    \kappa N-|\beta_{\rm m}|^2 
    =
    \frac{\kappa\big(N(N+1)-|M|^2\big)}{\Lambda},
\end{align}
Finally, for the terms multiplying the double commutators, we get
\begin{align}
    \frac{\kappa M^*}{2} + \frac{\alpha_{\rm m}\beta_{\rm m}^*}{2}
    =
    -\frac{\kappa\big(N(N+1)-|M|^2\big)}{2\Lambda}.
\end{align}
Likewise,
\begin{align}
    \frac{\kappa M}{2} + \frac{\alpha_{\rm m}^*\beta_{\rm m}}{2}
    =
    -\frac{\kappa\big(N(N+1)-|M|^2\big)}{2\Lambda}.
\end{align}
Therefore, we can conclude that the unmonitored superoperator has the form
\begin{align}
    \mathcal L_{\rm u}\rho
    =
    \Gamma \left(
    \mathcal D[\hat{c}]\rho
    +
    \mathcal D[\hat{c}^\dagger]\rho
    -
    \frac{1}{2}[\hat{c},[\hat{c},\rho]]
    -
    \frac{1}{2}[\hat{c}^\dagger,[\hat{c}^\dagger,\rho]]
    \right)\,,
\end{align}
with 
\begin{equation}
    \Gamma = \frac{\kappa\big(N(N+1)-|M|^2\big)}{\Lambda}.
\end{equation}
By defining the unmonitored operator 
\begin{align}
    \hat{A}_u = \sqrt{\Gamma} (\hat{c} + \hat{c}^\dagger) \,,
    \label{eq:Au}
\end{align}
we can use the identity in Eq.~\eqref{eq:identity} to prove that
\begin{align}
    \mathcal D[\hat{A}_{u}]\rho
    =
    \Gamma
    \left(
    \mathcal D[\hat{c}]\rho
    +
    \mathcal D[\hat{c}^\dagger]\rho
    -
    \frac{1}{2}[\hat{c},[\hat{c},\rho]]
    -
    \frac{1}{2}[\hat{c}^\dagger,[\hat{c}^\dagger,\rho]]
    \right).
\end{align}
Thus, the unmonitored superoperator corresponds to a Lindbladian 
\begin{align}
    \mathcal{L}_u\rho = \mathcal D[\hat{A}_{u}]\rho \,.
\end{align}

We can now finally state our main result, that is the equivalent form of the 
homodyne SME for a squeezed thermal bath as 
\begin{align}
    d\rho_c &= \mathcal{D}[\hat{A}_u]\rho_c \,dt + \mathcal{D}[\hat{A}_m]\rho_c \,dt \nonumber \\
    &\,\,\,\,\,\,\, + \mathcal{H}[\hat{A}_m]\rho_c \,dw_t \,.
    \label{eq:Ame}
\end{align}
Remarkably, we have thus demonstrated that the deterministic (unconditional) part of the SME is still written in the Lindblad form, but in terms of an unmonitored operator $\hat{A}_u$, defined in Eq.~\eqref{eq:Au}, and of a monitored operator $\hat{A}_m$, defined in Eq.~\eqref{eq:Am}, which in fact is the only one entering the measurement backaction stochastic part.

The identity
\begin{equation}
    N(N+1) - |M|^2 = \bar n_{\rm th}(\bar n_{\rm th}+1)
\end{equation}
implies that for a pure squeezed-vacuum environment ($\nth=0$), $\hat A_u = 0$. In this scenario, one thus obtains as expected
 a single collapse (monitored) operator
\begin{equation}
    \hat{A}_m
    = \sqrt{\kappa} \frac{\cosh r+e^{-i\theta}\sinh r}{\sqrt{\Lambda}}\hat b.
\end{equation}
 It follows from the identity
 \begin{equation}
    \Lambda = (2 \bar n_{\rm th} + 1) |\cosh r+e^{-i\theta}\sinh r|^2,
    \label{eq:Lambda_id}
\end{equation}
that, when $\bar n_{\rm th} = 0$, $\hat A_m$ differs from $\hat d_1$ by only a phase factor. This results in an effective homodyne phase shift.

\subsection{Interpretation in terms of unitary invariance of the Lindblad master equation}

The equivalence between the unconditional master equation in Eq.~\eqref{eq:me_v2} and the deterministic part of Eq.~\eqref{eq:Ame} can be understood as a consequence of the unitary freedom in the choice of Lindblad jump operators,
\begin{align}
\sum_{j=1}^n \mathcal{D}[\hat{c}_j]\rho = \sum_{j=1}^n \mathcal{D}\left[ \sum_{k=1}^n U_{jk} \hat{c}_k \right]\rho\,
\end{align}
with $U$ denoting a unitary matrix.

In our case, the two collapse operators appearing in Eq.~\eqref{eq:me_v2} are
\begin{equation}
    \hat d_1=\sqrt{\kappa(\bar n_{\rm th}+1)}\,\hat b,
    \qquad
    \hat d_2=\sqrt{\kappa\bar n_{\rm th}}\,\hat b^\dagger.
\end{equation}
Using the inverse Bogoliubov transformation,
\begin{equation}
    \hat c=\cosh r\,\hat b+e^{i\theta}\sinh r\,\hat b^\dagger,
\end{equation}
we can rewrite
\begin{align}
    \hat A_m
    &=
    \nu\sqrt{\frac{\bar n_{\rm th}+1}{\Lambda}}
    \hat d_1
    -
    \nu^*\sqrt{\frac{\bar n_{\rm th}}{\Lambda}}
    \hat d_2,
    \label{eq:Am_d}
    \\
    \hat A_u
    &=
    \nu\sqrt{\frac{\bar n_{\rm th}}{\Lambda}}
    \hat d_1
    +
    \nu^*\sqrt{\frac{\bar n_{\rm th}+1}{\Lambda}}
    \hat d_2,
\end{align}
with $\nu = \cosh r+e^{-i\theta}\sinh r$.
Equivalently,
\begin{equation}
    \begin{pmatrix}
        \hat A_m\\
        \hat A_u
    \end{pmatrix}
    =
    U
    \begin{pmatrix}
        \hat d_1\\
        \hat d_2
    \end{pmatrix},
\end{equation}
with
\begin{equation}
    U=
    \frac{1}{\sqrt{\Lambda}}
    \begin{pmatrix}
        \nu\sqrt{\bar n_{\rm th}+1}
        &
        -\nu^*\sqrt{\bar n_{\rm th}}
        \\[1mm]
        \nu\sqrt{\bar n_{\rm th}}
        &
        \nu^*\sqrt{\bar n_{\rm th}+1}
    \end{pmatrix}.
\end{equation}
\subsection{Evolution in terms of Kraus-like operators}
We consider a diffusive homodyne SME in the form 
\begin{align}
   d\rho_c &= -i[\hat{H},\rho_{c}(t)]dt +\sum_{j=1}^K\mathcal{D}[\hat{L}_j] \rho_c(t) dt \nonumber \\
   &\,\, +\sum_{j=1}^K\sqrt{\eta_j} \mathcal{H}[\hat{L}_j] \rho_c(t) \,dw_t^{(j)} \,,
   \label{eq:sme_gen}
\end{align}
corresponding to homodyne monitoring of $K$ photocurrents $dy_t^{(j)}=\sqrt{\eta_j} \langle \hat{L}_j + \hat{L}_j^\dagger\rangle \,dt + dw_t^{(j)}$. The efficiencies satisfy $0\leq \eta_j\leq 1$, such that $\eta_j = 0$ and $\eta_j=1$ correspond respectively to an unmonitored and perfectly monitored channel. 
The conditional evolution can be expressed in terms of Kraus-like operators via the formula \cite{Amini2011,ROUCHON20222,Jordan2016} 
\begin{equation}
    \rho_c(t+dt) =
    \frac{\tilde{\rho}_c(t+dt)}{\Tr[\tilde{\rho}_c(t+dt)]}\,,
    \label{eq:evcond}
\end{equation}
where the unnormalised state is
\begin{equation}
    \begin{split}
    \tilde{\rho}_c(t+dt) &=
    \hat{M}_{{\bf dy}_t}\rho_c(t)
    \hat{M}_{{\bf dy}_t}^{\dagger} \\
    &\quad+ \sum_j(1-\eta_j)
    \hat{L}_j\rho_c(t)
    \hat{L}_j^{\dagger}\,dt.
    \end{split}
\end{equation}
The Kraus-like operator is
\begin{align}
    \hat{M}_{{\bf dy}_t} = \mathbb{I} -i \hat{H} dt - \sum_{j}\left( \frac{1}{2} \hat{L}_j^\dagger\hat{L}_j dt - \sqrt{\eta_j}\, \hat{L}_j\, dy_t^{(j)}\right) \,,
    \label{eq:Kraus}
\end{align}
with ${\bf dy}_t = \{dy_t^{(1)},...,dy_t^{(K)}\}$ denoting the vector of the outcomes of the $K$ measurement channels. The evolution of the conditional state according to Eq.~\eqref{eq:evcond} is ensured to be both completely-positive and trace-preserving.

Since the SME we have derived in Eq.~\eqref{eq:Ame} has the same form of the general SME~\eqref{eq:sme_gen} with a perfectly monitored channel $\hat{A}_m$ and an unmonitored channel $\hat{A}_u$, we can now apply the formula in Eq.~\eqref{eq:evcond} for the evolved conditional state obtaining
\begin{align}
    \rho_c(t+dt) = \frac{\hat{M}_{{dy}_t} \rho_c(t) \hat{M}_{{dy}_t}^\dag +  \hat{A}_u \rho_c(t) \hat{A}_u^\dag \,dt }{\Tr[\hat{M}_{{dy}_t} \rho_c(t) \hat{M}_{{dy}_t}^\dag + \hat{A}_u \rho_c(t) \hat{A}_u^\dag \,dt]} \,,
    \label{eq:Kraus_evo}
\end{align}
where in this case the Kraus-like operator reads
\begin{align}
    \hat{M}_{dy_t} = \mathbb{I} -  \frac{1}{2}\left(\hat{A}_m^\dagger\hat{A}_m + \hat{A}_u^\dagger \hat{A}_u\right) dt + \hat{A}_m\, dy_t \,,
\end{align}
and with the photocurrent $dy_t$ defined in Eq.~\eqref{eq:photocurrent}.

Following the same recipe, one can also apply formulas for the evolved conditional state that take into account higher-order corrections, as described in~\cite{Rouchon2015,Guevara2020,Wonglakhon2024}, or the experimentally relevant scenario of time-averaged continuous measurements~\cite{robinet,wonglakhon2026,hutin_robinet_2026}.\\

We finally remark that the clear distinction between monitored and unmonitored channels also enables the use of existing routines for simulating diffusive SMEs, such as the one developed in the QuTiP package~\cite{JOHANSSON20121760,JOHANSSON20131234}.
\section{Frequency estimation via continuous homodyne detection in a squeezed-thermal bath}
\label{s:frequency_estimation}
Let us consider the estimation of the frequency $\omega$ for a classically driven two-level atom, whose emitted radiation in a squeezed-thermal bath is monitored via homodyne detection. The evolution of the conditional state is described by the SME
\begin{align}
    d\rho_c &= -i \omega [\hat{\sigma}_y,\rho_c]\,dt + \mathcal{L}\rho_c \,dt + \mathcal{H}[\hat{A}_m] \rho_c \,dw_t \,,
\end{align}
where the Lindbladian superoperator $\mathcal{L}$ and the monitored operator $\hat{A}_m$ are obtained respectively via the definitions in Eqs.~\eqref{eq:me_v1} and~\eqref{eq:Am} by fixing $\hat{c}=\hat{\sigma}_-$. For the numerical simulations, we initialize the qubit in the pure state corresponding to the eigenvalue $-1$ of $\sigma_z$. For simplicity we consider a {\em real squeezing parameter}, that is $\theta=0$ and thus $M\in \mathbb{R}$, such that the continuous photocurrent increment in Eq.~\eqref{eq:photocurrent} simply reads
\begin{align}
    dy_t 
         &= e^{-r}\sqrt{\frac{\kappa}{2\bar{n}_{th}+1}}  \langle \hat{\sigma}_x \rangle \,dt + dw_t \label{eq:photocurrent_qubit}
\end{align}
Since the deterministic component of the photocurrent is proportional to the expectation value of $\hat\sigma_x$, it naturally encodes information about the rotation frequency $\omega$ around the $y$-axis. 

\begin{figure*}[t]
    \centering
    \includegraphics[width=0.98\linewidth]{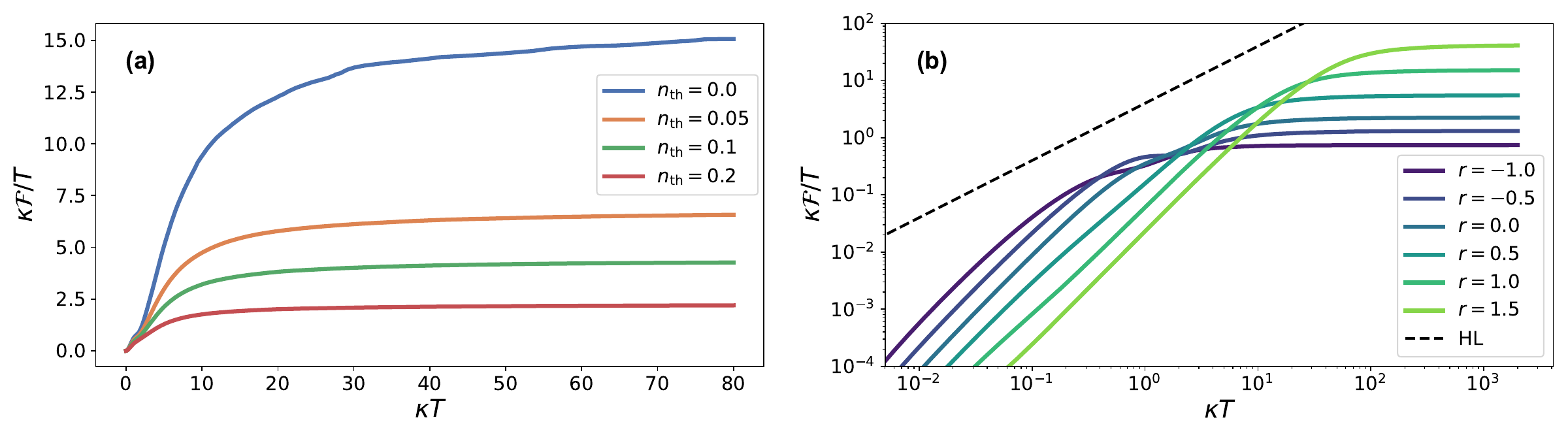}
    \caption{The time-averaged homodyne Fisher information rate, $\mathcal{F}_{\rm{hom}}/T$ of the frequency of a qubit coupled to a thermal environment for {\bf (a)}: different values of the thermal occupation at zero squeezing ($r = 0$); and {\bf (b)}: fixed thermal occupation ($\bar n_{\rm th} = 0.2$) with different values of the squeezing parameter $r$. The black dashed line indicates the Heisenberg limit (HL), $\mathcal{F}_{\rm{hom}}/T=4T$, which bounds the achievable Fisher information rate for Hamiltonian frequency estimation. The results were obtained by averaging over $N_{\rm traj} = 10^5$ trajectories in (a) and $N_{\rm traj} = 2\times 10^4$ trajectories in (b). In all simulations, $\omega/\kappa = 0.5$ and the qubit is prepared in the ground state of $\sigma_z$.}
    \label{f:figure}
\end{figure*}

\par

By making use of the decomposition into monitored and unmonitored jump operators derived in Sec.~\ref{s:equivalent_SME}, we can evaluate the Fisher information (FI) $\mathcal{F}_{\rm hom}$ in the homodyne signal by applying the Kraus-like evolution introduced in Eq.~\eqref{eq:Kraus_evo} via the method described in~\cite{AlbarelliQuantum}. As detailed in Appendix~\ref{a:FI_calc}, we adapt this approach to remove explicit zero-mean stochastic contributions to the Fisher information estimate.
This is particularly effective in numerical calculations as it reduces noise at long interrogation times, and also provides analytical insight leading to the following formula for the signal FI (see Appendix~\ref{a:FI_calc} for details)
\begin{equation}
    \mathcal{F}_{\rm hom} 
    = 
    \frac{\kappa e^{-2r}}{2 \bar n_{\rm th} + 1} \int _0^T dt \, \mathbb{E} \left [ (\Tr(\sigma_x \partial_\omega \rho_t))^2\right]
\end{equation}
This highlights that the information contained in the measurement record is determined not only by the amplitude of the homodyne signal, but also by the sensitivity of the conditional state to the unknown parameter, as we will indeed observe in the following.

We have first considered the thermal bath scenario without squeezing ($r=0$) and we show the results in Fig.~\ref{f:figure}\textbf{(a)}: $\mathcal{F}_{\rm hom}$ is plotted as a function of time and for different values of the number of thermal excitations $N=\bar{n}_{th}$. We observe that, as expected, the Fisher information decreases with increasing $\bar{n}_{th}$. 

We now investigate the effect of environmental squeezing at fixed, non-zero thermal occupation in Fig.~\ref{f:figure}\textbf{(b)}. The FI rate displays two distinct regimes. At short times, it is suppressed as $e^{-2r}$ as we might expect from Eq.~\eqref{eq:photocurrent_qubit}: positive (negative) values of $r$ suppress (amplify) the homodyne signal and thus, in principle, also the information on the frequency $\omega$. The FI grows superlinearly as $\mathcal{F}_{\rm hom} \propto T^3$. This scaling cannot continue forever because of the fundamental Heisenberg limit for Hamiltonian parameter estimation problems~\cite{boixo2007generalized}
\begin{equation}
    \mathcal{F}_{\rm hom}(T) 
    \leq 
    \|\partial_\omega H \|^2T^2 \leq 4T^2.
\end{equation}
Here $\|A\| = \lambda_{\max} - \lambda_{\min}$ is the spectral diameter of $A$. The resulting crossover occurs on a timescale $t \sim e^{2r}$, this can be understood from the relaxation rate derived in Appendix~\ref{a:gQFI}. After this time, we enter a linear scaling regime, and we can extract a steady-state rate. 

Remarkably, despite the suppression of the signal at early times, the long-time FI rate displays advantageous $e^{2r}$ scaling. Importantly, this enhancement persists at finite temperature, despite the corresponding dynamics containing an unmonitored dissipative channel.

Similar crossovers in the FI rate have been investigated in critical dissipative systems~\cite{Ilias2022} and in boundary time crystals~\cite{lee2025timescales}. Here we consider a mixed squeezed bath and restrict access to its physical homodyne output.

To help elucidate the numerical results, we analytically calculate the steady-state global QFI rate~\cite{GammelmarkPRL} in Appendix~\ref{a:gQFI}.
The global QFI $\mathcal{F}_{\rm global}$ sets the ultimate bound on the estimation precision by optimizing over all the (possibly joint) measurements on system and environmental degrees of freedom.
At steady-state, it can be evaluated via the formula~\cite{Ilias2022,oconnor2026attaining}
\begin{equation}
    \frac{\mathcal{F}_{\rm global}}{T} 
    = 
    8 \Re \int_0^\infty d\tau \, \,\Tr[e^{\mathcal{L}^\dag\tau} [\delta \hat{O}] (\delta \hat{O} \rho_{ss})]
\end{equation}
with $\delta\hat{O} = \partial_\omega H - \langle\partial_\omega H\rangle_{ss}$. In our example $\delta \hat{O} = \hat \sigma_y$ and we obtain
\begin{equation}
    \frac{\mathcal{F}_{\rm global}}{T}
    =
    \frac{16}{\kappa(2 \bar n_{\rm th}+1)}e^{2r}.
\end{equation}
This agrees with our non-zero temperature numerical results; temperature negatively affects the global QFI whereas environmental squeezing exponentially enhances it.
Note that the global QFI is derived assuming a pure global system-environment state~\cite{GammelmarkPRL}.
Thus, this QFI actually refers to the purified squeezed-thermal environment discussed in Sec.~\ref{s:squeezed_thermal_ME_SME}, rather than the information contained in the output modes of the thermal environment.

Finally, we investigate why the amplification of the homodyne signal for negative squeezing does not translate into an improvement in estimation precision. To show this, we calculate the long-time fluctuations of the integrated current~\cite{landi2024current}
\begin{align}
    Q_t &= \int_0^t dy_t \\
    D &= \lim_{t \to \infty} \frac{{\rm Var}(Q_t)}{t}.
\end{align}
In Appendix~\ref{a:gQFI} we show that for sufficiently large negative squeezing, exponential $e^{-r}$ amplification of the signal also results in exponential amplification of the noise as $D \sim e^{-4r}$.
The noise overpowers the signal and suppresses the steady-state rate. These results highlight the importance of analysing the full measurement record, rather than its amplitude alone, in determining the achievable sensitivity.

\section{Conclusions}
\label{s:conclusions}
We have derived an explicit monitored-unmonitored Lindblad decomposition of the homodyne SME for a squeezed-thermal environment.
By identifying the monitored and unmonitored channels, we have shown how to write the corresponding evolution in terms of Kraus-like operators. We use these to numerically evaluate the Fisher information for frequency estimation for a two-level atom emitting photons in a squeezed-thermal bath and continuously monitored via homodyne detection.
This method can also be applied to general-dyne unravellings of squeezed-thermal master equations~\cite{GeneralDino}.
Regardless of the physical implementation of such detection schemes, formally the corresponding SME can always be written as the one obtained by splitting the Lindblad generator in two terms and separately unravelling each term with homodyne measurements corresponding to two orthogonal quadratures~\cite{GeneralDino}.

Our numerical results for frequency estimation via a qubit in a squeezed-thermal bath show that, although temperature generally suppresses sensitivity, exponential squeezing enhancement of the steady-state Fisher information rate persists at finite temperature, despite the presence of unmonitored dissipation. This decomposition makes the calculation straightforward using existing numerical methods.

From a broader perspective, the approach presented in this paper provides an equivalence between the conditional dynamics of two physical scenarios.
In the original model, a quantum system is driven by an input (broadband) squeezed-thermal field, with the output field continuously monitored via homodyne detection.
In the equivalent vacuum-input model, the same quantum system is driven by two vacuum input fields through suitable jump operators, one monitored via homodyne and one unmonitored.
We stress that the equivalence holds at the level of conditional dynamics, i.e., the observed photocurrent and corresponding conditional states, not for the underlying quantum state of the output fields.
This is important to keep in mind, since the split into monitored and unmonitored channels may, for example, suggest to apply the methods of Refs.~\cite{Khan2025,Yang2026a} to evaluate or lower-bound the QFI of the output field.
However, those methods are derived under the assumption of vacuum input fields, thus they would evaluate the QFI of the output field for the second equivalent physical model described above, not for the original squeezed-thermal input field.

We expect that our results will be relevant in assessing the role of finite temperature for continuously monitored superconducting qubits in circuit-QED experiments when residual thermal microwave photons cannot be neglected~\cite{Yan2018,Wang2019} and in squeezing-enhanced qubit readout protocols~\cite{Eddins2018,Dassonneville2021,Li2026b}.

\section*{Acknowledgments}
The authors acknowledge useful discussions with Howard Wiseman. This research was supported in part by the International Centre for Theoretical Sciences (ICTS) for the program Quantum Trajectories (code: ICTS/qt2025/01).

\appendix

\section{Steady-state global quantum Fisher information and homodyne-current fluctuations.}
\label{a:gQFI}
We start by evaluating the steady-state global QFI rate of the classically driven two-level atom introduced in Sec~\ref{s:frequency_estimation}. We then investigate the fluctuations in the photocurrent and demonstrate that enhancing the signal amplitude alone is not sufficient for increased sensing precision.

For a mixing Lindbladian where the unknown parameter is encoded in the system Hamiltonian, the steady-state global QFI rate takes the form of a stationary two-time correlation function~\cite{Ilias2022,oconnor2026attaining}
\begin{align}
    \frac{\mathcal{F}_{\rm global}}{T} = 8 \Re \int_0^\infty d\tau \, \,\Tr[e^{\mathcal{L}^\dag\tau} [\delta \hat{O}] (\delta \hat{O} \rho_{ss})]
    \label{eq:global_qfi}
\end{align}
where $\delta\hat{O} = \partial_\omega H - \langle\partial_\omega H\rangle_{ss}$. In our example $\delta\hat{O} = \hat{\sigma}_y$, this commutes with the Hamiltonian, the remaining part to consider is the dissipators. For real squeezing ($\theta = 0$), the jump operators $A_m$ and $A_u$ both take the form $\hat{c} = a \hat\sigma_- + b \hat\sigma_+$ with real $a$ and $b$. It is useful to rewrite this in the form
\begin{equation}
    a \hat\sigma_- + b \hat\sigma_+ = \frac{a+b}{2}\hat\sigma_x - i\frac{a-b}{2} \hat\sigma_y.
\end{equation}
Using the Pauli commutation relations we find
\begin{equation}
    \mathcal{D}^\dag[a \hat\sigma_- + b \hat\sigma_+] \sigma_y = -\frac{(a+b)^2}{2} \sigma_y.
\end{equation}
For the two jump operators, this gives
\begin{align}
    \mathcal{D}^\dag[A_m] \sigma_y &= -\frac{(\alpha_m + \beta_m)^2}{2} \sigma_y = -\frac{\kappa}{2\Lambda} \sigma_y \\
    \mathcal{D}^\dag[A_u] \sigma_y &= -2\Gamma \sigma_y = -\frac{2\kappa \bar n_{\rm th}(\bar n_{\rm th}+1)}{\Lambda}\sigma_y.
\end{align}
Summing these together and using Eq.~\eqref{eq:Lambda_id} we arrive at the result
\begin{equation}
    \mathcal{L}^\dag[\sigma_y] = -\frac{\kappa}{2} (2 \bar n_{\rm th} + 1) e^{-2r} \sigma_y.
    \label{eq:gap}
\end{equation}
This allows the correlation-function integral in Eq.~\eqref{eq:global_qfi} to be evaluated analytically, giving the global QFI rate
\begin{equation}
    \frac{\mathcal{F}_{\rm global}}{T}
    =
    \frac{16}{\kappa(2 \bar n_{\rm th}+1)}e^{2r}.
\end{equation}
This demonstrates that the global QFI rate increases exponentially with positive squeezing, while thermal excitations suppress it. Note that this is the global QFI associated with the purified environment rather than the true thermal environment.

This result may come as a surprise given the photocurrent derived in Eq.~\eqref{eq:photocurrent_qubit}. The signal in the photocurrent is amplified with respect to the white noise term when $r<0$ and attenuated when $r>0$. However, Eq.~\eqref{eq:gap} tells us that the suppressed decay rate is the more important factor for the global QFI, which displays the opposite scaling with $r$. For continuously monitored signals, a more appropriate quantification of signal noise is the variance of the accumulated current~\cite{landi2024current}
\begin{align}
    Q_t &= \int_0^t dy_t \\
    D &= \lim_{t \to \infty} \frac{{\rm Var}(Q_t)}{t}.
\end{align}
The steady-state signal noise can be evaluated using the quantum regression theorem~\cite{landi2024current}
\begin{equation}
    D = 1 + 2\int_0^\infty \left[\Tr\left(\mathcal{J} e^{\mathcal{L}}[\mathcal{J} \rho_{ss}]  \right) - \Tr\left( \mathcal{J} \rho_{ss}\right)^2\right]
\end{equation}
with
\begin{equation}
    \mathcal{J}X = \hat A_m X + X \hat A_m^\dag
\end{equation}

\begin{figure}[t]
    \centering
    \includegraphics[width=0.98\linewidth]{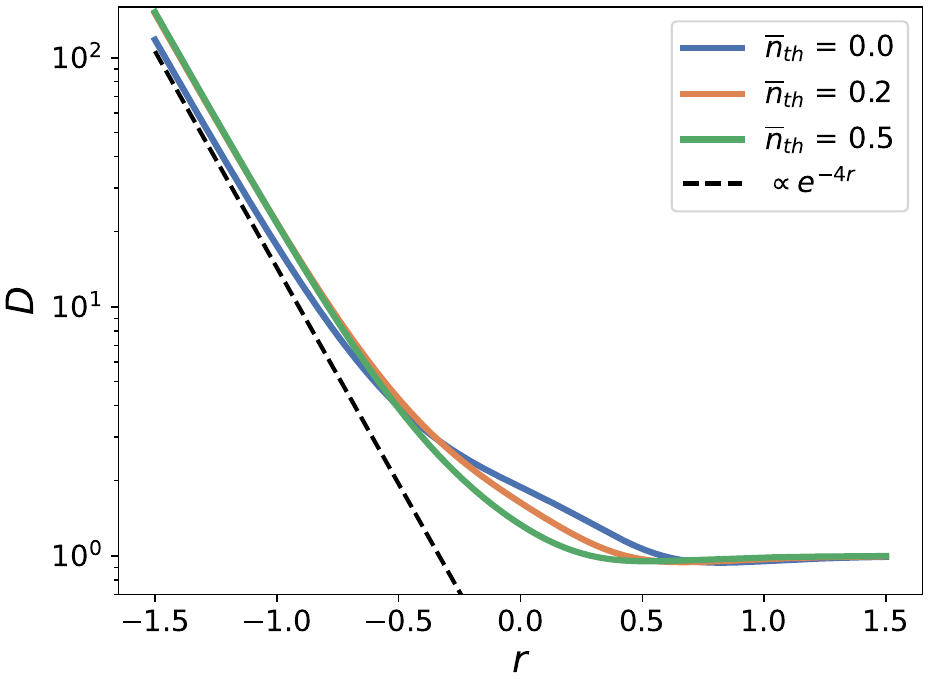}
    \caption{The steady-state signal noise, $D$ as a function of bath squeezing $r$ at various temperatures. \textit{Parameters}: $\omega/\kappa = 0.5$ }
    \label{f:noise}
\end{figure}

In Fig.~\ref{f:noise} we plot the steady-state signal noise against bath squeezing. For sufficiently large negative squeezing, the noise scales as $e^{-4r}$, removing any advantage from the $e^{-2r}$ amplification of the squared signal. For large positive squeezing, the noise is suppressed and tends to the white noise baseline.

\section{Calculation of the time-dependent signal Fisher information}
\label{a:FI_calc}

In this appendix, we derive an alternative expression for the Fisher information contained in a continuous homodyne measurement record. Our task is to estimate the unknown parameter $\omega$ from the measurement record $Y_{[0,t]}$. The Fisher information of the measurement data is
\begin{equation}
    \mathcal{F}_{\rm hom} = \mathbb E_\omega\left[
    \left(\partial_\omega\log p_{\rm{traj}}^{(t)}(Y_{[0,t]}|\omega)\right)^2
    \right].
\end{equation}
Directly calculating the conditional probability is difficult for long measurement times as it becomes exponentially small. Instead of calculating the conditional probability directly, one can evolve the unnormalized conditional state $\widetilde\varrho_t$. The trace of this operator is the conditional probability of that particular trajectory.
Gammelmark and Mølmer~\cite{GammelmarkMolmer} pointed out that multiplication of the state by a parameter-independent factor does not change the resulting score. This insight led to the introduction of the so-called monitoring operator~\cite{AlbarelliQuantum, radaelli2026parameter}
\begin{equation}
    \tau_t
    =
    \frac{\partial_\omega\widetilde\varrho_t}
         {\Tr\widetilde\varrho_t},
    \qquad
    S_t 
    = 
    \partial_\omega\log p_{\mathrm{traj}}^{(t)}
    =
    \Tr\tau_t
    .
\end{equation}
The repeated rescaling of the conditional derivative solves the problem of the operator trace vanishing at long measurement times. The signal FI is then usually estimated by averaging over many trajectories
\begin{equation}
    \mathcal{F}_{\rm hom} = \mathbb E[S_t^2].
\end{equation}
However, as we will show, this includes additional zero-mean stochastic terms that we can analytically remove. For simplicity, we restrict to the case of a single monitored operator, but the result directly generalises by summing the contributions from each operator. We note that an equivalent expression for the Fisher information of continuously monitored linear Gaussian quantum systems was derived in Ref.~\cite{genoni2017cramer}. Here we derive the corresponding result directly from the likelihood of the continuous homodyne measurement record, without assuming Gaussian conditional dynamics. In particular, this formulation can be easily implemented by adapting the numerical approach and code in Ref.~\cite{AlbarelliQuantum}.

Write the measurement current increment in the form
\begin{equation}
    dy_t = h_t dt + dw_t.
\end{equation}
The conditional probability distribution of the increment conditioned on the measurement history is a normal distribution centered at $h_t dt$ with variance $dt$
\begin{equation}
    p_\omega(dy_t|Y_{[0,t]}) = \frac{1}{\sqrt{2 \pi dt}} \exp \left[ - \frac{(dy_t - h_t dt)^2}{2 dt}\right].
\end{equation}
We calculate the derivative of the log-likelihood holding the observed increment, $dy_t$, and history, $Y_{[0,t]}$, fixed
\begin{equation}
    \partial_\omega \log p_\omega(dy_t|Y_{[0,t]}) = \partial_\omega h_t(dy_t - h_t dt) = \partial_\omega h_t \, dw_t.
\end{equation}
Therefore, the increment of the score is
\begin{equation}
    dS_t = \partial_\omega h_t \, dw_t.
\end{equation}
This allows us to calculate the FI increment using Itô calculus
\begin{align}
    d \mathcal{F}_{\rm hom} &= \mathbb E[d (S_t^2)] \nonumber \\
    &= \mathbb E [2 S_t d S_t+ (dS_t)^2] \nonumber \\
    &= \mathbb E[(\partial_\omega h_t)^2] dt.
\end{align}
giving the integrated FI
\begin{equation}
    \mathcal{F}_{\rm hom} = \int_0^t ds \, \mathbb E[(\partial_\omega h_s)^2].
\end{equation}
Using Itô's formula, the original approach gives
\begin{equation}
    S_t^2=
    \int_0^t ds \,(\partial_\omega h_s)^2
    +
    2\int_0^t dw_s \, S_s(\partial_\omega h_s).
\end{equation}
Therefore, this new approach explicitly eliminates a zero-mean stochastic contribution to the FI estimate. 

The final step is the calculation of $\partial_\omega h_t$: for an operator $\hat{c}$ monitored at efficiency $\eta$ this takes the form
\begin{equation}
    \partial_\omega h_t = \sqrt{\eta} \Tr[(\hat{c} + \hat{c}^\dag) \partial_\omega \rho_t] + 2 \sqrt{\eta} \Re \Tr[\partial_\omega\hat{c} \, \rho_t].
\end{equation}
Because of the relation~\cite{AlbarelliQuantum}
\begin{equation}
    \partial_\omega \rho_t = \tau_t - \rho_t \Tr \tau_t
\end{equation}
one can adapt the standard Kraus operator methods for the evolution of $\rho_t$ and $\tau_t$~\cite{AlbarelliQuantum} directly to this new method.

For the example considered in Sec.~\ref{s:frequency_estimation}, $h_t$ can be read from Eq.~\eqref{eq:photocurrent_qubit}. Since $\hat A_m$ is independent of $\omega$ the Fisher information takes the form
\begin{equation}
    \mathcal{F}_{\rm hom} = \frac{\kappa e^{-2r}}{2 \bar n_{\rm th} + 1} \int _0^T dt \, \mathbb{E} \left [ (\Tr(\sigma_x \partial_\omega \rho_t))^2\right]
\end{equation}
This highlights that the information contained in the measurement record is determined not only by the amplitude of the homodyne signal, but also by the sensitivity of the conditional state to the unknown parameter. Although positive squeezing exponentially suppresses the instantaneous signal, its effect on the conditional dynamics can lead to an enhanced Fisher information rate, as demonstrated numerically in Sec.~\ref{s:frequency_estimation}. 

Finally, we mention that this approach is closely related to established methods in classical statistics, where the Fisher information can similarly be calculated from the quadratic variation of the likelihood score~\cite{Barndorff-Nielsen1991,Barndorff-Nielsen1994,Surace2019}.

\bibliography{references_tidy}

@article{Ilias2022,
  title = {{{Criticality-Enhanced Quantum Sensing}} via {{Continuous Measurement}}},
  author = {Ilias, Theodoros and Yang, Dayou and Huelga, Susana F. and Plenio, Martin B.},
  year = 2022,
  month = mar,
  journal = {PRX Quantum},
  volume = {3},
  number = {1},
  eprint = {2108.06349},
  pages = {010354},
  publisher = {American Physical Society},
  doi = {10.1103/PRXQuantum.3.010354},
  url = {https://arxiv.org/abs/2108.06349},
  archiveprefix = {arXiv}
}

@article{GammelmarkPRL,
  title = {Fisher {{Information}} and the {{Quantum Cram{\'e}r-Rao Sensitivity Limit}} of {{Continuous Measurements}}},
  author = {Gammelmark, S{\o}ren and M{\o}lmer, Klaus},
  year = 2014,
  month = apr,
  journal = {Phys. Rev. Lett.},
  volume = {112},
  number = {17},
  eprint = {1310.5802},
  pages = {170401},
  publisher = {American Physical Society},
  doi = {10.1103/PhysRevLett.112.170401},
  url = {https://arxiv.org/abs/1310.5802},
  archiveprefix = {arXiv}
}

@article{Didier2015,
  title = {Fast {{Quantum Nondemolition Readout}} by {{Parametric Modulation}} of {{Longitudinal Qubit-Oscillator Interaction}}},
  author = {Didier, Nicolas and Bourassa, J{\'e}r{\^o}me and Blais, Alexandre},
  year = 2015,
  month = nov,
  journal = {Phys. Rev. Lett.},
  volume = {115},
  number = {20},
  eprint = {1504.04002},
  pages = {203601},
  publisher = {American Physical Society},
  doi = {10.1103/PhysRevLett.115.203601},
  url = {https://arxiv.org/abs/1504.04002},
  archiveprefix = {arXiv}
}

@article{Eddins2018,
  title = {Stroboscopic {{Qubit Measurement}} with {{Squeezed Illumination}}},
  author = {Eddins, Andrew and Schreppler, Sydney and Toyli, David M. and Martin, Leigh S. and {Hacohen-Gourgy}, Shay and Govia, Luke C. G. and Ribeiro, Hugo and Clerk, Aashish A. and Siddiqi, Irfan},
  year = 2018,
  month = jan,
  journal = {Phys. Rev. Lett.},
  volume = {120},
  number = {4},
  eprint = {1708.01674},
  pages = {040505},
  publisher = {American Physical Society},
  doi = {10.1103/PhysRevLett.120.040505},
  url = {https://arxiv.org/abs/1708.01674},
  archiveprefix = {arXiv}
}

@article{Dassonneville2021,
  title = {Dissipative {{Stabilization}} of {{Squeezing Beyond}} 3 {{dB}} in a {{Microwave Mode}}},
  author = {Dassonneville, R. and Assouly, R. and Peronnin, T. and Clerk, A. A. and Bienfait, A. and Huard, B.},
  year = 2021,
  month = may,
  journal = {PRX Quantum},
  volume = {2},
  number = {2},
  eprint = {2102.02863},
  pages = {020323},
  publisher = {American Physical Society},
  doi = {10.1103/PRXQuantum.2.020323},
  url = {https://arxiv.org/abs/2102.02863},
  archiveprefix = {arXiv}
}

@article{Mundarain2006,
  title = {Zeno and anti-{{Zeno}} effect for a two-level system in a squeezed bath},
  author = {Mundarain, D. F. and Stephany, J.},
  year = 2006,
  month = apr,
  journal = {Phys. Rev. A},
  volume = {73},
  number = {4},
  eprint = {quant-ph/0510214},
  pages = {042113},
  publisher = {American Physical Society},
  doi = {10.1103/PhysRevA.73.042113},
  url = {https://arxiv.org/abs/quant-ph/0510214},
  archiveprefix = {arXiv}
}

@article{Jakob2001,
  title = {Quantum measurement apparatus with a squeezed reservoir: {{Control}} of decoherence and nonlocality in phase space},
  author = {Jakob, Matthias and Abranyos, Yonatan and Bergou, J{\'a}nos A.},
  year = 2001,
  month = nov,
  journal = {Phys. Rev. A},
  volume = {64},
  number = {6},
  pages = {062102},
  publisher = {American Physical Society},
  doi = {10.1103/PhysRevA.64.062102},
  url = {https://link.aps.org/doi/10.1103/PhysRevA.64.062102}
}

@article{Cernotik2015,
  title = {Adiabatic elimination of {{Gaussian}} subsystems from quantum dynamics under continuous measurement},
  author = {{\v C}ernot{\'\i}k, Ond{\v r}ej and Vasilyev, Denis V. and Hammerer, Klemens},
  year = 2015,
  month = jul,
  journal = {Phys. Rev. A},
  volume = {92},
  number = {1},
  eprint = {1503.07484},
  pages = {012124},
  publisher = {American Physical Society},
  doi = {10.1103/PhysRevA.92.012124},
  url = {https://arxiv.org/abs/1503.07484},
  archiveprefix = {arXiv}
}

@article{Wang2019,
  title = {Cavity {{Attenuators}} for {{Superconducting Qubits}}},
  author = {Wang, Z. and Shankar, S. and Minev, Z. K. and {Campagne-Ibarcq}, P. and Narla, A. and Devoret, M. H.},
  year = 2019,
  month = jan,
  journal = {Phys. Rev. Appl.},
  volume = {11},
  number = {1},
  eprint = {1807.04849},
  pages = {014031},
  publisher = {American Physical Society},
  doi = {10.1103/PhysRevApplied.11.014031},
  url = {https://arxiv.org/abs/1807.04849},
  archiveprefix = {arXiv}
}

@article{Yan2018,
  title = {Distinguishing {{Coherent}} and {{Thermal Photon Noise}} in a {{Circuit Quantum Electrodynamical System}}},
  author = {Yan, Fei and Campbell, Dan and Krantz, Philip and Kjaergaard, Morten and Kim, David K. and Yoder, Jonilyn L. and Hover, David and Sears, Adam and Kerman, Andrew J. and Orlando, Terry P. and Gustavsson, Simon and Oliver, William D.},
  year = 2018,
  month = jun,
  journal = {Phys. Rev. Lett.},
  volume = {120},
  number = {26},
  eprint = {1801.00467},
  pages = {260504},
  publisher = {American Physical Society},
  doi = {10.1103/PhysRevLett.120.260504},
  url = {https://arxiv.org/abs/1801.00467},
  archiveprefix = {arXiv}
}

@article{AlbarelliQuantum,
  title = {Restoring {{Heisenberg}} scaling in noisy quantum metrology by monitoring the environment},
  author = {Albarelli, Francesco and Rossi, Matteo A. C. and Tamascelli, Dario and Genoni, Marco G.},
  year = 2018,
  month = dec,
  journal = {Quantum},
  volume = {2},
  eprint = {1803.05891},
  pages = {110},
  publisher = {Verein zur F{\"o}rderung des Open Access Publizierens in den Quantenwissenschaften},
  doi = {10.22331/q-2018-12-03-110},
  url = {https://arxiv.org/abs/1803.05891},
  archiveprefix = {arXiv},
  author+an = {1=me}
}

@article{GammelmarkMolmer,
  title = {Bayesian parameter inference from continuously monitored quantum systems},
  author = {Gammelmark, S{\o}ren and M{\o}lmer, Klaus},
  year = 2013,
  month = mar,
  journal = {Phys. Rev. A},
  volume = {87},
  number = {3},
  eprint = {1212.5700},
  pages = {032115},
  publisher = {American Physical Society},
  doi = {10.1103/PhysRevA.87.032115},
  url = {https://arxiv.org/abs/1212.5700},
  archiveprefix = {arXiv}
}

@article{Ravotto,
  title = {Quantum filtering of a thermal master equation with a purified reservoir},
  author = {Genoni, Marco G. and Mancini, Stefano and Wiseman, Howard M. and Serafini, Alessio},
  year = 2014,
  month = dec,
  journal = {Phys. Rev. A},
  volume = {90},
  number = {6},
  eprint = {1411.1435},
  pages = {063826},
  publisher = {American Physical Society},
  doi = {10.1103/PhysRevA.90.063826},
  url = {https://arxiv.org/abs/1411.1435},
  archiveprefix = {arXiv}
}

@phdthesis{WisemanPhDThesis,
  title = {Quantum {{Trajectories}} and {{Feedback}}},
  author = {Wiseman, Howard M.},
  year = 1994,
  school = {University of Queensland}
}

@article{wieczorekOptimalStateEstimation2015,
  title = {Optimal {{State Estimation}} for {{Cavity Optomechanical Systems}}},
  author = {Wieczorek, Witlef and Hofer, Sebastian G. and {Hoelscher-Obermaier}, Jason and Riedinger, Ralf and Hammerer, Klemens and Aspelmeyer, Markus},
  year = 2015,
  month = jun,
  journal = {Phys. Rev. Lett.},
  volume = {114},
  number = {22},
  eprint = {1505.01060},
  pages = {223601},
  publisher = {American Physical Society},
  doi = {10.1103/PhysRevLett.114.223601},
  url = {https://arxiv.org/abs/1505.01060},
  urldate = {2023-07-08},
  archiveprefix = {arXiv}
}

@article{rossiObservingVerifyingQuantum2019,
  title = {Observing and {{Verifying}} the {{Quantum Trajectory}} of a {{Mechanical Resonator}}},
  author = {Rossi, Massimiliano and Mason, David and Chen, Junxin and Schliesser, Albert},
  year = 2019,
  month = oct,
  journal = {Phys. Rev. Lett.},
  volume = {123},
  number = {16},
  eprint = {1812.00928},
  pages = {163601},
  publisher = {American Physical Society},
  doi = {10.1103/PhysRevLett.123.163601},
  url = {https://arxiv.org/abs/1812.00928},
  urldate = {2023-07-08},
  archiveprefix = {arXiv}
}

@article{duan_concurrent_2025,
  title = {Concurrent spin squeezing and field tracking with machine learning},
  author = {Duan, Junlei and Hu, Zhiwei and Lu, Xingda and Xiao, Liantuan and Jia, Suotang and M{\o}lmer, Klaus and Xiao, Yanhong},
  year = 2025,
  month = jun,
  journal = {Nat. Phys.},
  volume = {21},
  number = {6},
  eprint = {2402.00536},
  pages = {909--915},
  publisher = {Nature Publishing Group},
  doi = {10.1038/s41567-025-02855-3},
  url = {https://arxiv.org/abs/2402.00536},
  urldate = {2025-08-17},
  archiveprefix = {arXiv}
}

@article{ficheux2018dynamics,
  title = {Dynamics of a qubit while simultaneously monitoring its relaxation and dephasing},
  author = {Ficheux, Quentin and Jezouin, Sebastien and Leghtas, Zaki and Huard, Benjamin},
  year = 2018,
  month = may,
  journal = {Nat. Commun.},
  volume = {9},
  number = {1},
  eprint = {1711.01208},
  pages = {1926},
  publisher = {Nature Publishing Group},
  doi = {10.1038/s41467-018-04372-9},
  url = {https://arxiv.org/abs/1711.01208},
  archiveprefix = {arXiv}
}

@article{Minev2019Catch,
  title = {To catch and reverse a quantum jump mid-flight},
  author = {Minev, Z. K. and Mundhada, S. O. and Shankar, S. and Reinhold, P. and {Guti{\'e}rrez-J{\'a}uregui}, R. and Schoelkopf, R. J. and Mirrahimi, M. and Carmichael, H. J. and Devoret, M. H.},
  year = 2019,
  month = jun,
  journal = {Nature},
  volume = {570},
  number = {7760},
  eprint = {1803.00545},
  pages = {200--204},
  publisher = {Nature Publishing Group},
  doi = {10.1038/s41586-019-1287-z},
  url = {https://arxiv.org/abs/1803.00545},
  archiveprefix = {arXiv}
}

@article{Murch2013Observing,
  title = {Observing single quantum trajectories of a superconducting quantum bit},
  author = {Murch, K. W. and Weber, S. J. and Macklin, C. and Siddiqi, I.},
  year = 2013,
  month = oct,
  journal = {Nature},
  volume = {502},
  number = {7470},
  eprint = {1305.7270},
  pages = {211--214},
  publisher = {Nature Publishing Group},
  doi = {10.1038/nature12539},
  url = {https://arxiv.org/abs/1305.7270},
  archiveprefix = {arXiv}
}

@article{CampagneIbarcq2016Observing,
  title = {Observing {{Quantum State Diffusion}} by {{Heterodyne Detection}} of {{Fluorescence}}},
  author = {{Campagne-Ibarcq}, P. and Six, P. and Bretheau, L. and Sarlette, A. and Mirrahimi, M. and Rouchon, P. and Huard, B.},
  year = 2016,
  month = jan,
  journal = {Phys. Rev. X},
  volume = {6},
  number = {1},
  eprint = {1511.01415},
  pages = {011002},
  publisher = {American Physical Society},
  doi = {10.1103/PhysRevX.6.011002},
  url = {https://arxiv.org/abs/1511.01415},
  archiveprefix = {arXiv}
}

@article{AlbarelliPLA2024,
  title = {A pedagogical introduction to continuously monitored quantum systems and measurement-based feedback},
  author = {Albarelli, Francesco and Genoni, Marco G.},
  year = 2024,
  month = jan,
  journal = {Phys. Lett. A},
  volume = {494},
  eprint = {2312.13214},
  pages = {129260},
  publisher = {Elsevier},
  doi = {10.1016/j.physleta.2023.129260},
  url = {https://arxiv.org/abs/2312.13214},
  urldate = {2023-12-21},
  archiveprefix = {arXiv}
}

@book{WisemanMilburn,
  title = {Quantum {{Measurement}} and {{Control}}},
  author = {Wiseman, Howard M. and Milburn, Gerard J.},
  year = 2010,
  publisher = {Cambridge University Press},
  address = {Cambridge},
  isbn = {1107424151},
  doi = {10.1017/CBO9780511813948}
}

@book{jacobs2014quantum,
  title = {Quantum {{Measurement Theory}} and its {{Applications}}},
  author = {Jacobs, Kurt},
  year = 2014,
  publisher = {Cambridge University Press},
  address = {Cambridge},
  doi = {10.1017/CBO9781139179027}
}

@book{JordanSiddiqi,
  title = {Quantum {{Measurement}}: {{Theory}} and {{Practice}}},
  author = {Jordan, Andrew N. and Siddiqi, Irfan A.},
  year = 2024,
  publisher = {Cambridge University Press},
  address = {Cambridge},
  doi = {10.1017/9781009103909}
}

@article{JOHANSSON20121760,
  title = {{{QuTiP}}: {{An}} open-source {{Python}} framework for the dynamics of open quantum systems},
  author = {Johansson, J. R. and Nation, P. D. and Nori, Franco},
  year = 2012,
  month = aug,
  journal = {Comput. Phys. Commun.},
  volume = {183},
  number = {8},
  eprint = {1110.0573},
  pages = {1760--1772},
  publisher = {Elsevier},
  doi = {10.1016/j.cpc.2012.02.021},
  url = {https://arxiv.org/abs/1110.0573},
  archiveprefix = {arXiv}
}

@article{JOHANSSON20131234,
  title = {{{QuTiP}} 2: {{A Python}} framework for the dynamics of open quantum systems},
  author = {Johansson, J. R. and Nation, P. D. and Nori, Franco},
  year = 2013,
  month = apr,
  journal = {Comput. Phys. Commun.},
  volume = {184},
  number = {4},
  eprint = {1211.6518},
  pages = {1234--1240},
  publisher = {Elsevier},
  doi = {10.1016/j.cpc.2012.11.019},
  url = {https://arxiv.org/abs/1211.6518},
  archiveprefix = {arXiv}
}

@article{wonglakhon2026,
  title = {Quantum trajectories for time-binned data and their closeness to fully conditioned quantum trajectories},
  author = {Wonglakhon, Nattaphong and Chantasri, Areeya and Wiseman, Howard M.},
  year = 2026,
  month = oct,
  journal = {Phys. Rev. A},
  volume = {114},
  number = {4},
  eprint = {2601.10937},
  pages = {042415},
  publisher = {American Physical Society},
  doi = {10.1103/9gbb-7lk6},
  url = {https://arxiv.org/abs/2601.10937},
  archiveprefix = {arXiv}
}

@misc{hutin_robinet_2026,
  title = {Calculus of {{Robinet}}: completely positive reconstruction of time-averaged diffusive quantum trajectories},
  author = {Hutin, Hector and Tilloy, Antoine},
  year = 2026,
  month = jul,
  number = {arXiv:2607.18392},
  eprint = {2607.18392},
  publisher = {arXiv},
  doi = {10.48550/arXiv.2607.18392},
  url = {https://arxiv.org/abs/2607.18392},
  archiveprefix = {arXiv}
}

@article{Jordan2016,
  title = {Anatomy of fluorescence: quantum trajectory statistics from continuously measuring spontaneous emission},
  author = {Jordan, Andrew N. and Chantasri, Areeya and Rouchon, Pierre and Huard, Benjamin},
  year = 2016,
  month = sep,
  journal = {Quantum Stud.: Math. Found.},
  volume = {3},
  number = {3},
  eprint = {1511.06677},
  pages = {237--263},
  publisher = {Springer},
  doi = {10.1007/s40509-016-0075-9},
  url = {https://arxiv.org/abs/1511.06677},
  archiveprefix = {arXiv}
}

@misc{robinet,
  title = {Time-averaged continuous quantum measurement},
  author = {Guilmin, Pierre and Rouchon, Pierre and Tilloy, Antoine},
  year = 2025,
  month = may,
  number = {arXiv:2505.20382},
  eprint = {2505.20382},
  publisher = {arXiv},
  doi = {10.48550/arXiv.2505.20382},
  url = {https://arxiv.org/abs/2505.20382},
  archiveprefix = {arXiv}
}

@article{GeneralDino,
  title = {General-dyne unravelling of a thermal master equation},
  author = {Genoni, Marco G. and Mancini, Stefano and Serafini, Alessio},
  year = 2014,
  month = jul,
  journal = {Russ. J. Math. Phys.},
  volume = {21},
  number = {3},
  eprint = {1405.3565},
  pages = {329--336},
  publisher = {Pleiades Publishing},
  doi = {10.1134/S1061920814030054},
  url = {https://arxiv.org/abs/1405.3565},
  archiveprefix = {arXiv}
}

@article{Wonglakhon2024,
  title = {Completely positive trace-preserving maps for higher-order unraveling of {{Lindblad}} master equations},
  author = {Wonglakhon, Nattaphong and Wiseman, Howard M. and Chantasri, Areeya},
  year = 2024,
  month = dec,
  journal = {Phys. Rev. A},
  volume = {110},
  number = {6},
  eprint = {2408.14105},
  pages = {062207},
  publisher = {American Physical Society},
  doi = {10.1103/PhysRevA.110.062207},
  url = {https://arxiv.org/abs/2408.14105},
  archiveprefix = {arXiv}
}

@article{Guevara2020,
  title = {Completely positive quantum trajectories with applications to quantum state smoothing},
  author = {Guevara, Ivonne and Wiseman, Howard M.},
  year = 2020,
  month = nov,
  journal = {Phys. Rev. A},
  volume = {102},
  number = {5},
  eprint = {1909.12455},
  pages = {052217},
  publisher = {American Physical Society},
  doi = {10.1103/PhysRevA.102.052217},
  url = {https://arxiv.org/abs/1909.12455},
  archiveprefix = {arXiv}
}

@article{Rouchon2015,
  title = {Efficient quantum filtering for quantum feedback control},
  author = {Rouchon, Pierre and Ralph, Jason F.},
  year = 2015,
  month = jan,
  journal = {Phys. Rev. A},
  volume = {91},
  number = {1},
  eprint = {1410.5345},
  pages = {012118},
  publisher = {American Physical Society},
  doi = {10.1103/PhysRevA.91.012118},
  url = {https://arxiv.org/abs/1410.5345},
  archiveprefix = {arXiv}
}

@inproceedings{Amini2011,
  title = {On stability of continuous-time quantum filters},
  author = {Amini, Hadis and Mirrahimi, Mazyar and Rouchon, Pierre},
  year = 2011,
  month = dec,
  booktitle = {Proc. 50th IEEE Conf. Decis. Control Eur. Control Conf. (CDC-ECC)},
  eprint = {1103.2706},
  pages = {6242--6247},
  publisher = {IEEE},
  address = {Orlando, FL, USA},
  doi = {10.1109/CDC.2011.6160631},
  url = {https://arxiv.org/abs/1103.2706},
  archiveprefix = {arXiv}
}

@article{ROUCHON20222,
  title = {A tutorial introduction to quantum stochastic master equations based on the qubit/photon system},
  author = {Rouchon, Pierre},
  year = 2022,
  journal = {Annu. Rev. Control},
  volume = {54},
  eprint = {2208.07416},
  pages = {252--261},
  publisher = {Elsevier},
  doi = {10.1016/j.arcontrol.2022.09.006},
  url = {https://arxiv.org/abs/2208.07416},
  archiveprefix = {arXiv}
}

@article{Yang2026a,
  title = {Quantum {{Cram{\'e}r-Rao Precision Limit}} of {{Noisy Continuous Sensing}}},
  author = {Yang, Dayou and Ketkar, Moulik and Audenaert, Koenraad and Huelga, Susana F. and Plenio, Martin B.},
  year = 2026,
  month = feb,
  journal = {Phys. Rev. Lett.},
  volume = {136},
  number = {7},
  eprint = {2504.12400},
  pages = {070802},
  publisher = {American Physical Society},
  doi = {10.1103/ltfw-4fwn},
  url = {https://arxiv.org/abs/2504.12400},
  urldate = {2026-02-20},
  archiveprefix = {arXiv}
}

@article{Khan2025,
  title = {Tensor {{Network Approach}} to {{Sensing Quantum Light-Matter Interactions}}},
  author = {Khan, Aiman and Albarelli, Francesco and Datta, Animesh},
  year = 2025,
  month = nov,
  journal = {PRX Quantum},
  volume = {6},
  number = {4},
  eprint = {2504.12399},
  pages = {040343},
  publisher = {American Physical Society},
  doi = {10.1103/ljh3-3l4j},
  url = {https://arxiv.org/abs/2504.12399},
  urldate = {2025-11-21},
  archiveprefix = {arXiv}
}

@article{oconnor2026attaining,
  title = {Attaining quantum sensing enhancement from monitored dissipative time crystals},
  author = {O'Connor, Eoin and Montenegro, Victor and Albarelli, Francesco and Paris, Matteo G. A. and Bayat, Abolfazl and Genoni, Marco G.},
  year = 2026,
  month = sep,
  journal = {npj Quantum Inf.},
  eprint = {2508.15448},
  publisher = {Nature Publishing Group},
  doi = {10.1038/s41534-026-01331-x},
  url = {https://arxiv.org/abs/2508.15448},
  archiveprefix = {arXiv}
}

@article{landi2024current,
  title = {Current {{Fluctuations}} in {{Open Quantum Systems}}: {{Bridging}} the {{Gap Between Quantum Continuous Measurements}} and {{Full Counting Statistics}}},
  author = {Landi, Gabriel T. and Kewming, Michael J. and Mitchison, Mark T. and Potts, Patrick P.},
  year = 2024,
  month = apr,
  journal = {PRX Quantum},
  volume = {5},
  number = {2},
  eprint = {2303.04270},
  pages = {020201},
  publisher = {American Physical Society},
  doi = {10.1103/PRXQuantum.5.020201},
  url = {https://arxiv.org/abs/2303.04270},
  archiveprefix = {arXiv}
}

@article{radaelli2026parameter,
  title = {Parameter estimation for quantum jump unraveling},
  author = {Radaelli, Marco and Smiga, Joseph A. and Landi, Gabriel T. and Binder, Felix C.},
  year = 2026,
  month = feb,
  journal = {Quantum},
  volume = {10},
  eprint = {2402.06556},
  pages = {1993},
  publisher = {Verein zur F{\"o}rderung des Open Access Publizierens in den Quantenwissenschaften},
  doi = {10.22331/q-2026-02-02-1993},
  url = {https://arxiv.org/abs/2402.06556},
  archiveprefix = {arXiv}
}

@article{genoni2017cramer,
  title = {{{Cram{\'e}r-Rao}} bound for time-continuous measurements in linear {{Gaussian}} quantum systems},
  author = {Genoni, Marco G.},
  year = 2017,
  month = jan,
  journal = {Phys. Rev. A},
  volume = {95},
  number = {1},
  eprint = {1608.08429},
  pages = {012116},
  publisher = {American Physical Society},
  doi = {10.1103/PhysRevA.95.012116},
  url = {https://arxiv.org/abs/1608.08429},
  archiveprefix = {arXiv}
}

@article{Barndorff-Nielsen1991,
  title = {Information {{Quantities}} in {{Non-Classical Settings}}},
  author = {{Barndorff-Nielsen}, O. E. and S{\o}rensen, M.},
  year = 1991,
  month = sep,
  journal = {Comput. Stat. Data Anal.},
  volume = {12},
  number = {2},
  pages = {143--158},
  publisher = {Elsevier},
  doi = {10.1016/0167-9473(91)90014-S},
  url = {https://www.sciencedirect.com/science/article/pii/016794739190014S},
  urldate = {2026-10-18}
}

@article{Barndorff-Nielsen1994,
  title = {A {{Review}} of {{Some Aspects}} of {{Asymptotic Likelihood Theory}} for {{Stochastic Processes}}},
  author = {{Barndorff-Nielsen}, O. E. and S{\o}rensen, M.},
  year = 1994,
  journal = {Int. Stat. Rev.},
  volume = {62},
  number = {1},
  eprint = {1403550},
  eprinttype = {jstor},
  pages = {133--165},
  publisher = {Wiley, International Statistical Institute (ISI)},
  doi = {10.2307/1403550},
  url = {https://www.jstor.org/stable/1403550},
  urldate = {2026-10-18}
}

@article{Surace2019,
  title = {Online {{Maximum-Likelihood Estimation}} of the {{Parameters}} of {{Partially Observed Diffusion Processes}}},
  author = {Surace, Simone Carlo and Pfister, Jean-Pascal},
  year = 2019,
  month = jul,
  journal = {IEEE Trans. Autom. Control},
  volume = {64},
  number = {7},
  eprint = {1611.00170},
  pages = {2814--2829},
  publisher = {IEEE},
  doi = {10.1109/TAC.2018.2880404},
  url = {https://arxiv.org/abs/1611.00170},
  urldate = {2026-10-18},
  archiveprefix = {arXiv}
}

@article{boixo2007generalized,
  title = {Generalized {{Limits}} for {{Single-Parameter Quantum Estimation}}},
  author = {Boixo, Sergio and Flammia, Steven T. and Caves, Carlton M. and Geremia, {\relax JM}},
  year = 2007,
  month = feb,
  journal = {Phys. Rev. Lett.},
  volume = {98},
  number = {9},
  eprint = {quant-ph/0609179},
  pages = {090401},
  publisher = {American Physical Society},
  doi = {10.1103/PhysRevLett.98.090401},
  url = {https://arxiv.org/abs/quant-ph/0609179},
  urldate = {2026-10-22},
  archiveprefix = {arXiv}
}

@misc{lee2025timescales,
  title = {Timescales, {{Squeezing}} and {{Heisenberg Scalings}} in {{Many-Body Continuous Sensing}}},
  author = {Lee, Gideon and Belyansky, Ron and Jiang, Liang and Clerk, Aashish A.},
  year = 2025,
  month = may,
  number = {arXiv:2505.04591},
  eprint = {2505.04591},
  publisher = {arXiv},
  doi = {10.48550/arXiv.2505.04591},
  url = {https://arxiv.org/abs/2505.04591},
  urldate = {2025-05-08},
  archiveprefix = {arXiv}
}

@article{Li2026b,
  title = {Full-{{Counting Statistics}} and {{Quantum Information}} of {{Dispersive Readout}} with a {{Squeezed Environment}}},
  author = {Li, Ming and Luo, JunYan and Platero, Gloria and Engelhardt, Georg},
  year = 2026,
  month = apr,
  journal = {Phys. Rev. Lett.},
  volume = {136},
  number = {14},
  eprint = {2512.02531},
  pages = {143601},
  publisher = {American Physical Society},
  doi = {10.1103/2s1m-y9bd},
  url = {https://arxiv.org/abs/2512.02531},
  urldate = {2026-04-06},
  archiveprefix = {arXiv}
}

@article{Murch2013a,
  title = {Reduction of the Radiative Decay of Atomic Coherence in Squeezed Vacuum},
  author = {Murch, K. W. and Weber, S. J. and Beck, K. M. and Ginossar, E. and Siddiqi, I.},
  year = 2013,
  month = jul,
  journal = {Nature},
  volume = {499},
  number = {7456},
  eprint = {1301.6276},
  pages = {62--65},
  publisher = {Nature Publishing Group},
  doi = {10.1038/nature12264},
  url = {https://arxiv.org/abs/1301.6276},
  urldate = {2025-02-01},
  archiveprefix = {arXiv}
}

@article{Toyli2016,
  title = {Resonance {{Fluorescence}} from an {{Artificial Atom}} in {{Squeezed Vacuum}}},
  author = {Toyli, D. M. and Eddins, A. W. and Boutin, S. and Puri, S. and Hover, D. and Bolkhovsky, V. and Oliver, W. D. and Blais, A. and Siddiqi, I.},
  year = 2016,
  month = jul,
  journal = {Phys. Rev. X},
  volume = {6},
  number = {3},
  eprint = {1602.03240},
  pages = {031004},
  publisher = {American Physical Society},
  doi = {10.1103/PhysRevX.6.031004},
  url = {https://arxiv.org/abs/1602.03240},
  urldate = {2026-10-26},
  archiveprefix = {arXiv}
}

@article{Gardiner1986,
  title = {Inhibition of {{Atomic Phase Decays}} by {{Squeezed Light}}: {{A Direct Effect}} of {{Squeezing}}},
  author = {Gardiner, Crispin W.},
  year = 1986,
  month = may,
  journal = {Phys. Rev. Lett.},
  volume = {56},
  number = {18},
  pages = {1917--1920},
  doi = {10.1103/PhysRevLett.56.1917},
  url = {https://link.aps.org/doi/10.1103/PhysRevLett.56.1917}
}

@article{Nurdin2014,
  title = {Quantum Filtering for Multiple Input Multiple Output Systems Driven by Arbitrary Zero-Mean Jointly {{Gaussian}} Input Fields},
  author = {Nurdin, Hendra I.},
  year = 2014,
  month = jul,
  journal = {Russ. J. Math. Phys.},
  volume = {21},
  number = {3},
  eprint = {1409.2666},
  pages = {386--398},
  doi = {10.1134/S106192081403011X},
  url = {https://arxiv.org/abs/1409.2666},
  archiveprefix = {arXiv}
}

@misc{Gough2026a,
  title = {Quantum {{Filtering}} for {{Squeezed Noise Inputs}}},
  author = {Gough, John and Rees, Dylon},
  year = 2026,
  month = jan,
  number = {arXiv:2601.12564},
  eprint = {2601.12564},
  publisher = {arXiv},
  doi = {10.48550/arXiv.2601.12564},
  url = {http://arxiv.org/abs/2601.12564},
  urldate = {2026-01-21},
  archiveprefix = {arXiv}
}

\end{document}